# Search engines in polarized media environment: Auditing political information curation on Google and Bing prior to 2024 US elections

**Mykola Makhortykh\*, Tobias Rorhbach**[1]\*, Maryna Sydorova\*, and Elizaveta Kuznetsova\*\*

\*Institute of Communication and Media Studies, University of Bern, Fabrikstrasse 8, 3012 Bern, Switzerland

\*\*Research group "Platform Algorithms and Digital Propaganda", Weizenbaum Institute, Hardenbergstraße 32, 10623 Berlin, Germany

**Declaration of Interest statement**

The authors have no competing interests that can be reported.

**Acknowledgments**

Data collections for the research were sponsored by the Friends of the Institute of Communication and Media Science at the University of Bern grant "Auditing how the performance of web search engines regarding representation of epistemically contested political issues changes over time" awarded to Mykola Makhortykh and Maryna Sydorova and the Swiss National Science Foundation grant "Algorithm audit of the impact of user- and system-side factors on web search bias in the context of federal popular votes in Switzerland" (PI: Mykola Makhortykh; 105217_215021).

---

[1] The authors, whose names are bolded, have contributed equally to the study design and implementation and share the first authorship.

# Search engines in polarized media environment: Auditing political information curation on Google and Bing prior to 2024 US elections

**Abstract**: Search engines play an important role in the context of modern elections. By curating information in response to user queries, search engines influence how individuals are informed about election-related developments and perceive the media environment in which elections take place. It has particular implications for (perceived) polarization, especially if search engines' curation results in a skewed treatment of information sources based on their political leaning. Until now, however, it is unclear whether such a partisan gap emerges through information curation on search engines and what user- and system-side factors affect it. To address this shortcoming, we audit the two largest Western search engines, Google and Bing, prior to the 2024 US presidential elections and examine how these search engines' organic search results and additional interface elements represent election-related information depending on the queries' slant, user location, and time when the search was conducted. Our findings indicate that both search engines tend to prioritize left-leaning media sources, with the exact scope of search results' ideological slant varying between Democrat- and Republican-focused queries. We also observe limited effects of location- and time-based factors on organic search results, whereas results for additional interface elements were more volatile over time and specific US states. Together, our observations highlight that search engines' information curation actively mirrors the partisan divides present in the US media environments and has the potential to contribute to (perceived) polarization within these environments.

Introduction

Search engines are key information curators in today's high-choice media environment. By selecting and ranking information sources in response to user queries, search engine algorithms help individuals worldwide navigate the abundance of information on topics ranging from health (Bragazzi et al., 2017) to historical atrocities (Makhortykh et al., 2022) to elections (Trielli & Diakopoulous, 2022). The ease of use and applicability for different types of information-seeking behavior, together with the perceived reliability of search outputs, make search engines highly trusted information sources (Schultheiß & Lewandowski, 2023). As a result, search engines are used by individuals to stay informed about important societal developments and play an important role in news exposure and discovery (Nielsen, 2016).

Due to their societal relevance, search engines increasingly come under scrutiny over the quality of information sources they expose their users to and the factors that affect such exposure[2]. Many of these studies focus on whether search engines' performance is prone to algorithmic bias, which can be defined as a tendency to produce outputs that are systematically skewed toward the viewpoints or interests of specific social groups (Friedman & Nissenbaum, 1996). While it can be debated to what degree search engines can be impartial, considering the existence of multiple forms of social bias and technological limitations in tackling them, engines' reiteration and amplification of biased representations of social reality can result in many societal harms. Examples of such harms vary from the propagation of discriminatory representations of social groups (Noble, 2018; Urman & Makhortykh, 2024) to the exposure of search engine users to false information (Kuznetsova et al., 2024).

---

[2] The importance of search engines has been reflected in the rapidly growing body of research on algorithmic information curation and its implications for information exposure. These studies differ in their methodological approaches, which range from algorithm audits (Unkel & Haim, 2021; Trielli & Diakopoulous, 2022; Urman et al., 2022a) to data donation- and webtracking-based studies (Blassnig

The risks associated with search engines are amplified by the growing societal polarization that characterizes many modern liberal democracies. While the exact scope of polarization remains debated by scholars, there exists growing evidence that politics are increasingly perceived as a highly polarising subject. Such a perception may be, to a large extent, shaped by the evolving political and media landscape that amplifies partisanship and contributes to "a self-perpetuating cycle" (Wilson et al., 2020: 223) of rising polarization. Considering that by design, search engine algorithms treat their users in a differentiated manner to better adapt search outputs to their needs (e.g., based on user language or location), it often leads to specific population groups potentially being exposed to different interpretations of the same societal issue (Makhortykh et al., 2024a). This makes search engines' information curation an important factor that can potentially amplify (perceived) societal polarization, in particular in already polarised media environments.

The relationship between search engines and polarization is of particular importance in the context of elections. Earlier studies have indicated that information curation by search engines can affect voter preferences by shifting their users' opinions about whom to vote for (Epstein & Robertson, 2015; Rohrbach et al., 2024). Under these circumstances, biased representation of election-related information, for instance, in the form of exposure of certain groups of voters to information sources with a particular ideological slant, can have implications for voting behavior. However, empirical research on search engines' information curation and its potential biases in the context of elections remains limited and often produces contradictory results (e.g., Diakopoulos et al., 2018; Unkel & Haim, 2021). Similarly, the effects of different factors affecting information curation (and its potential biases), such as user location of changes in source relevance, remain under-studied.

To address these gaps, we examine how the two largest Western search engines, Google and Bing, curated political information prior to the 2024 US elections. Our interest in the US context is due to it constituting one of the most polarised media environments among liberal democracies together with a particularly high degree of internet penetration and search engine use. Using virtual agent-based algorithm audits, we examine how the presence of specific types of political information sources is affected by user- (the use of ideologically slanted and non-slanted search queries) and system-side factors (user location and time-based changes in search relevance). Specifically, we focus on information curation of journalistic media with different political leaning in organic search results and other elements

---

et al., 2023; van Hoof et al., 2024) to the self-reported assessments (Zumofen, 2024; Vziatysheva et al., 2024).

of search interface (e.g., Newsblock for Google) due to journalistic media being the most common type of election-related search results.

The rest of the paper is organized as follows: first, we briefly review related work on the role of search engines in the context of polarization and in the time of elections, together with the impact of user- and system-side factors affecting their performance and formulate our research questions and a hypothesis. Then, we introduce the methodology used to conduct the algorithm audits and to analyze the collected data. After that, we present our findings regarding differences in the selection of sources and their ideological leaning depending on the slant of the query and the impact of search location and time-based changes. Finally, we conclude with a discussion of the implications of our findings, together with the limitations of the current study and directions for future research.

# Search engines, political information curation, and polarization

The way in which citizens in different countries inform themselves about politics is determined by media environments. These environments have been traditionally shaped by factors such as media regulation, journalistic culture, and the interplay of various media forms (Hoskins & Tulloch, 2016). With the rise of digital platforms and the expansion of the channels through which individuals are exposed to political information, media environments have undergone a series of fundamental changes. One particular change that we focus on in this paper concerns the increasing impact of algorithmic information curation systems on media environments and political information exposure within them. By "organizing, selecting and presenting subsets of a corpus of information" (Rader & Gray, 2015: 173), information curation systems influence how individuals are exposed to certain information sources and opinions. Examples of such systems range from content recommenders used by social media and legacy media to AI-powered chatbots such as Gemini to search engines such as Google and Bing.

The emergence of information curation systems has multiple implications for political information exposure. Such systems can enable non-intentional - or incidental (Thorson, 2020) - exposure to political information through (often non-transparent) curation decisions and facilitate selective exposure to content aligning with users' pre-existing beliefs. Aiming to guide users towards information that is relevant to their interests, curation systems customize their results to meet the information needs of individual users. In the case of political information, such customization may result in the reinforcement of pre-existing ideological beliefs and the isolation of users in personalized information bubbles (Borgesius et al., 2016). While empirical evidence of such bubbles remains inconclusive (Bruns, 2019), such a perspective is concerning in the context of democratic decision-making as it can undermine political participation and contribute to societal polarization.

The discussion of risks of information curation in the context of polarization has been largely focused on personalized content feeds on social media platforms (Yang et al., 2023) and automated news recommendations used by legacy media (Ludwig et al., 2023). The role of search engines in this process has been largely omitted from the discussion despite their extensive use around the world, together with growing evidence of algorithmic bias affecting how search engines curate political information (e.g., Pradel, 2021). Partially, such omittance can be attributed to the limited evidence of individual-based personalization of search engine results (e.g., Puschmann, 2019) as contrasted by more aggressive personalization used by other curation systems. However, less aggressive individual personalization does not fully exclude the possibility of search engines contributing to the skewed exposure to information, for instance, depending on the language that specific groups use for search queries or geography-based search localization. In turn, such skewness can contribute both to the divisive understanding of societal phenomena, which are represented by search engines, and a more polarized perception of media environments where search engines function.

Understanding how search engines shape media environments and individual perceptions of these environments is particularly important during political events such as elections. A number of studies show that search engines serve as essential entry points for information about politics (e.g., Dutton & Reisdorf, 2017). Considering the fundamental connection between informedness and citizens' ability to participate in political decision-making, information curation on search engines becomes an important factor in the electoral process due to it determining how individuals are exposed to information that can influence their voting decisions. Different forms of algorithmic bias associated with such curation can act as conduits of actual and perceived biases of users' media diet (Robertson et al., 2023), resulting in an even more polarised perception of electoral matters and their representation by media outlets. Consequently, by contributing to the (perceived) skewness of media environments, search engines can serve as an overlooked driver of polarization in the context of elections (for a review of this argument, see Kubin & Von Sikorski, 2023).

The importance of the relationship between search engines, politics, and polarization is reflected in a growing volume of studies that look at how search engines curate political information in the context of elections. These studies can be grouped into three main categories. The first category looks at the impact of information curation on voting behavior and tackles issues such as the search engine manipulation effect (Epstein & Robertson, 2015) and the effects of algorithmic bias on candidates' electability (Rohrbach et al., 2024). The second category focuses on the representation of political actors by search engines and whether such representation can be prone to different forms of bias (e.g., gender bias; Pradel, 2021; Rohrbach et al., 2024). Finally, the third category examines how search engines curate information in the context of elections, what factors affect such curation, and whether it results in the exposure of individuals to ideologically slanted sources (Robertson et al., 2023; Trielli & Diakopoulos, 2022; Urman et al., 2022a).

The latter group of studies is particularly relevant for understanding the implications of political information curation on search engines for polarised media environments. The findings of these studies, however, remain inconclusive. Some of them highlight the tendency of search engines to expose users to more mainstream information sources and interpretations (e.g., Unkel & Haim, 2021; Trielli & Diakopoulous, 2022), whereas others show that search engines curate information in a way that leads to the ideologically slanted selection of sources (Diakopoulous et al., 2018). One of the reasons for these divergent assessments can be that search engine outputs are shaped by multiple factors that are difficult to account for systematically. To address this challenge, we introduce a framework of

user- and system-side factors affecting search engine outputs and discuss how they can influence information curation on search engines in the context of elections.

## User-side factors of political information curation on search engines

A major challenge of studying search engines' political information curation relates to its highly complex and non-transparent nature (Castillo, 2019). To identify the most relevant set of information sources in response to user queries, search engine algorithms take into consideration many factors, including the presence of meta tags, source recency, the structure of URLs, and the mobile-friendliness of the source's website. Not only the composition and weight of factors determining individual search outputs is unclear, but these factors' importance varies between search engines. For instance, the ranking algorithm for organic search results on Google is assumed to give more preference to relevance and diversity of domains that link to the web page in question as well as authoritativeness of the domain associated with the webpage, whereas for Bing, more emphasis is placed on the content readability and freshness (Wilkinson, 2023).

While the above-mentioned factors play an important role in determining what information is retrieved by search engine algorithms, information curation is always dependent on users' input. Search engine results are, therefore, profoundly affected by a number of user-side factors, ranging from the topics of search queries and their exact formulations to the query language or the use of search operators such as "site:". Among these factors, most research to date focused on the impact of query selection on search engine outputs and identified strong variation in search results regarding topics ranging from migration (Norocel & Lewandowski, 2023) to conspiracy theories (Urman et al., 2022b) depending on queries' focus. Such variation causes the risk of individuals being exposed to fundamentally different interpretations of the same issue based on the queries' formulations, especially if search results are systematically skewed towards a specific viewpoint (e.g., anti-migration opinions for more right-leaning queries; Norocel & Lewandowski, 2023) and minimize exposure to alternative viewpoints. In turn, such skewed exposure can contribute to the perceived polarization of a media environment where the search is occurring.

In the case of election-related information curation, the discussion of the role of user-side factors has largely focused on how diverse or homogeneous is the selection of sources in response to queries with different thematic foci (Unkel & Haim, 2021; Trielli & Diakopoulous, 2022). The importance of this issue is attributed to evidence of search engines' information curation affecting voter preferences (Epstein & Robertson, 2015) and the possible variation in this effect based on different search queries. For instance, if for queries regarding certain candidates, search results are more likely to include sources promoting the position of the candidate (while not necessarily doing it for other candidates), it may result in different information exposure that, in turn, may have implications for voting behavior (e.g., due to different levels of electorate's mobilization). Such a risk has been amplified by the increasing use of search engines for political microtargeting (Dobber, 2023), which may further contribute to unequal information exposure.

In addition to affecting voting behavior, skewed exposure to information due to user-side factors can affect how search engine users perceive media environments in which the elections are happening and the degree of polarization therein. For instance, if an individual searches for information about the candidate they support and most results come from resources critical to the candidate, an individual can perceive it as evidence of media polarization, especially if such skewness is systematic. The degree to which search engine performance is skewed regarding election-related information, however, remains currently unclear. Trielli and Diakopoulos (2022) showed for the US context that despite differences in the query selection between Democratic and Republican voters, Google organic search results consist of a similar selection of sources that tends towards mainstream and centrist interpretations. Urman et al. (2022a), in a study of search results regarding US presidential candidates, found a strong homogeneity of Google search results, which were dominated by journalistic media. Similarly, Nechushtai et al. (2024) observed the tendency of Google to prioritize a similar selection of sources for queries with different ideological slant.

At the same time, other studies demonstrate that depending on the search query, search engines can prioritize rather different information sources, resulting in skewed exposure to election-related information. Urman et al. (2022a) observed that, unlike Google, Bing tends to return non-homogenous results for queries regarding individual candidates, including a higher presence of sources associated with candidates' campaigns for certain individuals. Unkel and Haim (2021), in their study on the German federal elections, noted variation in the selection of sources by Google depending on the search orientation of the users (e.g., interest in individual politicians or specific issues). While, according to the study, journalistic media prevailed in search results, queries associated with specific orientations resulted in fluctuations in the ranking of other types of sources, including the varying presence of specific parties' websites. Similarly, Schwabl et al. (2023), in their study of the 2021 federal elections in Germany, observed substantial variation in the selection of sources depending on the political party on which the query has been focusing.

At the time of writing (i.e., winter 2024-2025), there is a lack of studies regarding the role of search engines in the media environment during the 2024 US presidential elections. However, understanding this role is important, considering the high degree of (perceived) polarization that characterizes the US (Wilson et al., 2020) and the growing impact of societal polarization on elections worldwide, in particular regarding the growing appeal of populist parties that is often underpinned by polarization within specific societies. In the US case, there have been anecdotal observations that the representation of election-related information has been skewed in the case of search engines (e.g., due to higher visibility of left-leaning media; Kang, 2024), but such claims were not systematically tested. To assess the validity of these arguments and contribute to a better understanding of the impact of user-side factors on information curation in the context of elections, we investigate whether and how search queries with a partisan focus—that is, queries focusing on Republican or Democratic candidates (vs. neutral keywords—result in systematic differences regarding the types of sources prioritized in search outputs. As a first step, we formulate the following research question:

**RQ1**: What types of sources are prioritized by search engines during the 2024 US elections in response to Democratic-focused, Republican-focused, and neutral queries?

Secondly, we want to understand whether the selection of information sources by search engines reflects the ideological slant of the search query. Specifically, we focus here on the subset of data regarding search results coming from media sources, which we expect to constitute the majority of results based on earlier studies (e.g., Unkel & Haim, 2021; Urman et al., 2022) and for which it is easier to identify an ideological leaning (as contrasted with, for instance, most social media platforms). The prioritization of media with ideological slant aligning with the slant of a query (e.g., left-leaning media in the case of Democratic-focused queries) can contribute to the (perceived) polarization of a media environment during the 2024 elections. To test this expectation, we formulate the following hypothesis:

**H1**: Search engine results show (a) higher share of left-leaning media sources for Democratic-focused queries and (b) higher shares of right-leaning queries for Republican-focused queries.

## System-side factors of political information curation on search engines

Besides user-side factors, such as the choice of a search query, political information curation on search engines is shaped by the selection of system-side factors. Conceptually, system-side factors are factors outside the users' control that can be seen as moderating the outcome of information-seeking behavior. These are multiple system-side factors, including the randomization of results for the same queries, which is used for real-time A/B testing[3] on search engines (Makhortykh et al., 2020), the customization of search results based on the user location (Kliman-Silver et al., 2015) or results' personalization based on user search behavior (Hannak et al., 2013). Because it is hardly possible to account for the effect of all possible system-side factors, we specifically focus on the two prominent factors in information curation: user location and source relevance.

*Search localization* is one of the core functionalities of search engines which increases the relevance of search results by adapting them to the place from where the user is searching for information. A number of studies (e.g., Ballatore et al., 2017; Urman & Makhortykh, 2024; Kuznetsova et al., 2024) looked at the implications of search localization for representing different societal phenomena. Ballatore et al. (2017) demonstrated that users from higher-income countries are more likely to have localized sources compared with users from low-income countries, particularly regarding geography-related requests. Urman and Makhortykh (2024) showed that user location has a substantive impact on exposure to gender-discriminatory advertisements when searching for information about gender and national groups. Finally, Kuznetsova et al. (2024) found that user location is an important factor in determining exposure to information sources that promote disinformation regarding Russia's war in Ukraine.

Relatively little research, however, has been done on the effects of search location on election-related information, in particular the variation in the source selection on the regional (and not cross-country) level. Kliman-Silver et al. (2015) observed that queries dealing with politicians' names trigger relatively little localization, whereas queries referring to politics-related concepts result in more local results. Rohrbach et al. (2024) analyzed how queries in different languages used from different locations result in varying degrees of representation of female and male politicians in gender-neutral queries regarding local political institutions; however, their analysis neither focused on elections nor disentangled between the effects of the choice of a search query and effects of search localization. Finally, Perreault et al. (2024), in the analysis of Google search results in the context of the US 2022 elections, found not only the tendency of a search engine to prioritize local information sources for election-related information but also the frequent mistargeting (i.e., the incorrect matching between the user location and localized information sources).

---

[3] A/B testing is a form of controlled experiment, where individuals are exposed to one of two experimental conditions. In the context of search engines, A/B tests can be applied to compare the effects of source selection and ranking strategies on user engagement with search results.

While these findings show that the location may influence what type of information users see in response to their queries, they provide no guidance on how location is connected to a potential partisan gap in search engine outputs, which we understand as an unequal representation of information sources with different ideological leaning. However, such a connection can have significant implications for the (perceived) polarization of the media environment, especially user location determines exposure to sources with a specific ideological slant. One assumption would be that the partisan gap aligns with the political leaning of the location from which the search is conducted; that is, left-leaning sources will be more present when searching for election-related information in pro-Democratic states, while the visibility of right-leaning sources will be higher for the same queries in pro-Republican states. As this assumption lacks a clear empirical basis, we ask for the moderating influence of search location in the following research question:

**RQ2:** How does the search location affect the partisan gap between right- and left-leaning media sources in search engines' outputs?

Another system-side factor that we focus on is time-based changes in *source relevance*. This concept refers to the constant adaptation of search results to catch up with the emergence of more relevant information sources. Such an adaptation results in changes in the selection of search results over time, in particular for developing stories. On a daily basis, such changes tend to be relatively minor and usually result in re-ranking sources instead of adding new ones (Hannak et al., 2013); however, there is substantive variation between search engines, with Bing outputs being more volatile even during shorter periods of time (Kuznetsova et al., 2024). For longer periods of time, changes in source relevance can profoundly transform the representation of specific societal phenomena (Makhortykh et al., 2024b), which has implications for how these phenomena are perceived and interpreted.

In the context of political information, very few studies look at how source relevance changes over time. One of the exceptions is a study by Ulloa et al. (2024a) conducted during the 2020 US presidential elections. The results of the study demonstrated that search results regarding political information are prone to substantial time-based fluctuations, in particular around elections, and reiterated the observation by Kuznetsova et al. (2024) that Bing outputs are more volatile over time than Google outputs. The study also highlighted that source relevance is influenced by search localization, with the rate of changes in search results over time varying depending on the search location.

There are, however, some aspects of source relevance that remain unclear. One of them regards the implications of changes in source relevance for exposure to sources with different ideological slant. It is known that some sources (e.g., Wikipedia) are less prone to time-based changes and consistently remain in the top search results. However, the degree to which such consistency applies to media sources and how equally it applies to sources with specific ideological leaning remains unknown. The potential for differentiated treatment of sources based on their leaning is of particular relevance here due to its implications for perceived polarization. If users observe that right- or left-leaning sources are more likely to be consistently present in search results (whereas sources with an opposite leaning are more likely to disappear), then it may contribute to the polarized perception of a media environment. To address this gap, we formulate the following research question:

**RQ3**: How does the partisan gap between right- and left-leaning media sources in search engines' outputs evolve over the course of the election campaign?

# Methodology

To answer the research questions and test the hypothesis outlined above, we conducted an algorithm audit of two search engines, Google and Bing, in the run-up to the 2024 US presidential elections. Algorithm auditing is a research method used to systematically analyze the performance of complex decision-making systems and evaluate their functionality and impact (Mittelstadt, 2016). This method is often applied to study information curation and its potential biases in the case of search engines (e.g., Makhortykh et al., 2022; Urman et al., 2022a; Urman et al., 2022b; Kuznetsova et al., 2024) and other algorithmic systems, including news recommenders (Bandy & Diakopoulos, 2020; Daucé & Loveluck, 2021) or social media feeds (Bartley et al., 2021; Kuznetsova & Makhortykh, 2023).

There are different methodological approaches for algorithm audits (for reviews, see Bandy, 2021; Urman et al., 2024). However, most studies on search engines' political information curation use virtual agent-based audits that rely on programmatic simulation of human activity to generate inputs for the system (Ulloa et al., 2024). Usually, such audits involve simulating the process of entering queries into a search engine interface, scrolling down to load search results, and then saving HTML pages with results for subsequent processing. This approach enables more control over the data collection, including possibilities to control for the changes in source relevancy by collecting data at fixed time points and the effects of location by simulating user activity from specific IP addresses. Finally, virtual agent-based audits can be scaled by deploying multiple agents to simultaneously simulate the same set of actions, which can ensure that observations are not disproportionately affected by the randomization of search results (Makhortykh et al., 2020).

## Data collection

To collect data for the study, we developed a cloud-based infrastructure for deploying a large number of virtual agents hosted via Google Compute Engine. This cloud service also provides a broad range of IP addresses that can be used to simulate agent activity from specific regions. The infrastructure consisted of one storage server to preserve data during data collection and a large number of virtual machines (N=30 for monthly collections and N=60 for daily collections) with the same hardware and software specifications (with IP addresses distributed across the set of US states). Each virtual machine hosted two virtual agents using the Firefox browser; each agent was programmed to open the English version of Google (i.e., google.com), enter the search query, and then retrieve the first page of search results for Google text search for each search query (see below). To account for randomization, each search query consisting of the name of the candidate was simultaneously entered by 20 virtual agents per location (overall, 60 agents for monthly collections and 120 agents for daily collections). To minimize the impact of personalization, we closed the web browser after each query and cleaned cookies.

Our audits aimed to investigate the impact of user- and system-side factors on information curation regarding the US elections in order to evaluate how search engines may contribute to (perceived) polarization of the US media environment. To implement the audit, we used three types of election-related queries: (1) *neutral queries* without partisan cues ("elections united states", "elections president united states", "elections house united states", "elections congress united states"), (2) *Republican-focused queries* ("donald trump", "elections donald trump", "jd vance"), and (3) *Democratic-focused queries* ("joe biden", "elections joe biden", "kamala harris", "elections kamala harris", "tim walz"). For brevity, in the remaining part of the paper, we will refer to Republican- and Democratic-focused queries as Republican and Democratic, respectively. For Democratic queries, we replaced queries for Joe Biden with those for Kamala Harris after her candidacy was ratified by the Democratic convention.

To assess how search engines' curation is affected by time-based changes in source relevance, we repeated these queries in the six months prior to the election (June to October).[4] To capture short-term variability in the final phase of the election, we repeated the queries on a daily basis from 23 October until a few days after the election on 10 November.[5]

Every query was conducted in different US locations to investigate how information curation is affected by search localization. For monthly data collections, we ran the queries from IP addresses based in three regions: Las Vegas (NV), Los Angeles (CA), and Dallas (TX). For daily data collections, we used six locations: Las Vegas (NV), Los Angeles (CA), Salt Lake City (UT), Council Bluffs (IA), Columbus (OH), and Ashburn (VA). For every query, we collected the first page of search results for Google and Bing. Such a page contains between 7 and 10 organic search results, which are links to external information sources. For Google searches, we additionally distinguished between organic search results and Newsblock results, which were added a few months before the elections.

Google Newsblock is a news aggregator embedded in Google searches that indexes content from different media sources to provide users with a condensed overview of breaking developments on the issue that is searched for. We also treated as part of a Newsblock, a selection of results from social media platforms, which were sometimes added by Google below the news stories (usually, such results came from candidates' accounts on X). Due to its high visibility on the web page (i.e., before the organic search results), Newsblock is likely to attract significant attention from users, as indicated by earlier research on other elements of the search interface added by Google to expand organic search results. At the same time, due to being temporarily added to the search page results, Newsblock can follow different curation principles than Google's organic search results[6] and it can potentially lead to different patterns of political information exposure. However, there is little research on the role of Newsbllock (as well as other additional elements of the search interface) in algorithmic information curation due to most existing studies focusing only on organic search results.

---

[4] In July, we conducted a second ad hoc data collection after the assassination attempt on Donald Trump in Pennsylvania on 13 July.
[5] Due to a failure of the cloud-based data collection architecture, we are missing data for 27-29 October.

Overall, these data collections resulted in a total of 305'901 search results coming from Google and Bing (see Table 1 for an overview). To preprocess collected data, we first built customized HTML parsers to differentiate between content originating from Google's organic search results and Newsblock and to identify organic search results for Bing. We then parsed and extracted text results for all three types of search engines, where each search result constituted a domain of the information source (i.e., the URL that is linked), the title of the linked information source as well as metadata of the collection (i.e., query, location, date, identifier of the virtual agent).

**Table 1.** Overview of search result sample sizes for different search engine audits

|  | Query type | | |
| --- | --- | --- | --- |
| Data collection | Democrat | Neutral | Republican |
| Google - organic search | | | |
|     Monthly | 3'943 | 5'065 | 3'538 |
|     Daily | 19'314 | 32'648 | 21'864 |
| Google - Newsblock | | | |
|     Monthly | 5'018 | 3'698 | 3'702 |
|     Daily | 29'790 | 46'268 | 26'837 |
| Bing | | | |
|     Monthly | 4'130 | 4'485 | 3'538 |
|     Daily | 28'166 | 37'058 | 26'839 |
| Total search results | 90'361 | 129'222 | 86'318 |

*Note*. Numbers are aggregated across location and time of data collection (more detailed information on the different samples is provided in the supplemental materials).

---

[6] Specifically, Newsblock can be more driven by system-side factors due to its more volatile and situational nature.

## Measures

We first coded *source type* by extracting the root domains for each unique search result and classifying them as belonging to one of seven source categories: 1) journalistic (e.g., cnn.com, economist.com, usatoday.com, nytimes.com); 2) government (e.g., usa.gov, vote.gov, fec.gov); 3) social media (e.g., x.com, facebook.com, instagram.com); 4) NGO/think tanks (e.g., aclu.org, opensecrets.org, constitutioncenter.org); 5) candidate websites (e.g., kamalaharris.com, donaldjtrum.com, trump.com); 6) blogs (e.g., projects.fivethirtyeight.com, thecampaignworkshop.com); and 7) encyclopedias (e.g., wikipedia.org, ballotpedia.org). For the classification, we used existing lists of domains assigned to the above-mentioned categories from earlier auditing studies produced by the authors; to classify domains absent from these lists, we relied on manual labeling of source types conducted by one of the authors.

For the subset of journalistic sources, we determined the *political leaning* by matching each source to entries in the Media Bias Fact Check (MBFC) database.[7] The database contains over 7'800 sources whose political leaning is categorized from extreme left/right, left/right, left-/right-center, to least biased. Because the exact boundaries between the ideological subcategories are somewhat blurry, we recoded the categories into two dummy variables, indicating the presence of a political leaning to the left (1 = present; 0 = absent) or right (1 = present; 0 = absent).

Finally, we used the meta-data from the data collections as independent variables. The *query type* (neutral, Democratic, or Republican) constitutes the main user-side factor, whereas the *date* (month or day of the query) and the *location* (where the query was conducted) represent system-side factors.

## Data analysis

Data analysis consists of three steps. First, we provided a descriptive analysis of source types by aggregating the share of each source type within monthly and daily data collections (RQ1). Second, we focused on journalistic media, which constitute the most frequent category of search results, and ran generalized linear mixed-effects models with a logit link and random intercepts by a virtual agent to predict the likelihood of encountering a right- or left-leaning media source in search results (H1). The models include fixed effects for the query type as our main predictor of interest, controlling for date and location to account for their influence on the presence of slanted sources in search results. We ran separate models for left- and right-leaning sources and for the three types of search (organic Google text search, organic Bing text search, and Google Newsblock). Third, we assessed whether the distribution of left- and right-leaning sources varies across regions (RQ2) by repeating the models with by-region interaction effects and then computing Tukey-adjusted pairwise contrasts of the predicted probabilities (Democratic queries – Republican queries) for each region along with 95% confidence Intervals obtained with the Wald method. We then repeated the procedure by calculating the same contrasts for each time point for each data collection (RQ3). Full models and additional tables are provided in the Appendix.

---

[7] See https://mediabiasfactcheck.com/. Some examples of studies using the MBFC include Baly et al. (2018), Weld et al. (2021), Sánchez-Cortés et al. (2024).

# Results

## User-side factors of political information curation

We first investigate differences in the selection of sources in Google and Bing search results for Republican, Democratic, and neutral queries (RQ1). Panel A in Figure 1 depicts the shares of source types by query type aggregated across Bing and Google queries. For all query types, journalistic media sources constitute the largest shares in search result outputs, followed by encyclopedia and government sources (see Table B1 in the Appendix for a detailed breakdown).

Whereas the shares of most source types are rather similar for both Republican and Democratic queries, journalistic sources are more prevalent in the output of Republican queries by 14.1% in the monthly and 10.5% in daily searches. Panel B of Figure 1 shows that this difference in the share of journalistic media sources is rather stable over time, indicating a tendency for search engines to prioritize these sources for Republican queries. Interestingly, for Democratic queries, we observed a higher presence of government sources (which can be attributed to Democratic candidates being incumbent officials) as well as social media sources. Social media-related sources were almost absent from neutral queries, thus suggesting that for Democratic and Republican queries, search engines were likely to include links to candidates' social media accounts.

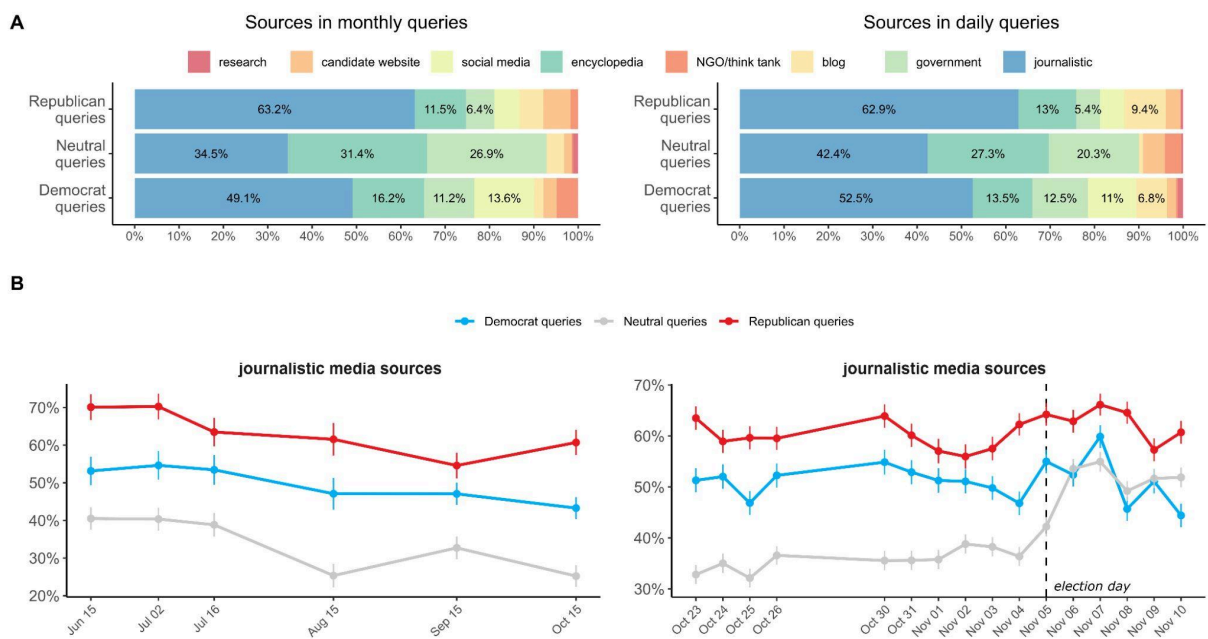

**Figure 1.** Descriptive overview of sources in search results for partisan and neutral queries. Panel A depicts mean shares of all sources by query type. Panel B depicts mean shares of journalistic media sources by query type across time, along with 95% confidence intervals. Data aggregated across organic Bing, organic Google and Google's Newsblock search results.

We then examine the political leaning of media sources to test whether the ideological slant of queries is reflected in slanted search outputs. Descriptively, we find that left-leaning media sources are substantially more present in the output of both Google and Bing, irrespective of the type of query, thus aligning with observations from Diakopoulous et al. (2018) regarding search engine outputs in the context of the 2016 US elections. Whereas roughly half of media sources have at least some leaning toward the political left, the share of right-leaning media sources only ranges between 15–35%. Among the ten most prevalent journalistic sources, only one source (www.foxnews.com) is considered leaning to the political right according to the MBFC database (see Tables B2.1–B2.2). Most other highly prevalent sources represent journalistic media outlets associated with a left-leaning slant (e.g., www.cnn.com, www.nytimes.com, www.washingtonpost.com, www.abcnews.com).

Although these *absolute* differences in the distribution of left- and right-leaning media sources are noteworthy, we were interested in whether search engine outputs also create the *relative* partisan gap (i.e., by prioritizing media sources with the leaning corresponding to the leaning of a search query). In line with H1a, we indeed find that Democratic queries are associated with a higher likelihood of left-leaning sources for the monthly data collections related to Bing ($OR$ = 1.23, $95\%CI$[1.10:1.37], $p$ < 0.001) and Google organic search results ($OR$ = 1.36, $95\%CI$[1.10:1.68], $p$ = 0.005) but not for Google's Newsblock outputs ($OR$ = 1.05, $95\%CI$[0.94:1.17], $p$ = 0.420). For instance, the predicted share of left-leaning sources in Bing search outputs is 61.2% ($95\%CI$[59.3:63.1%]) for Democratic queries but only 52.5% ($95\%CI$[50.1:54.4%]) for Republican queries (see panel A of Figure 2). For Bing but not Google, we also found that Republican queries significantly decrease the likelihood of left-leaning media appearing in search results ($OR$ = 0.88, $95\%CI$[0.79:0.98], $p$ = 0.022).

For daily collections, the pattern of search results is more mixed. On the one hand, we found the expected positive relationship between Democratic queries and left-leaning sources only for Bing ($OR$ = 1.17, $95\%CI$[1.12:1.22], $p$ < 0.001). For organic and Newsblock outputs on Google, this effect is not statistically significant. On the other hand, there is consistent evidence that Republican queries decrease the likelihood of left-leaning sources for all three types of search (Bing: $OR$ = 0.94, $95\%CI$[0.90:0.97], $p$ = 0.002; Google-organic: $OR$ = 0.73, $95\%CI$[0.69:0.78], $p$ < 0.001; Google-Newsblock: $OR$ = 0.77, $CI95\%$[0.74 – 0.80], $p$ < 0.001).

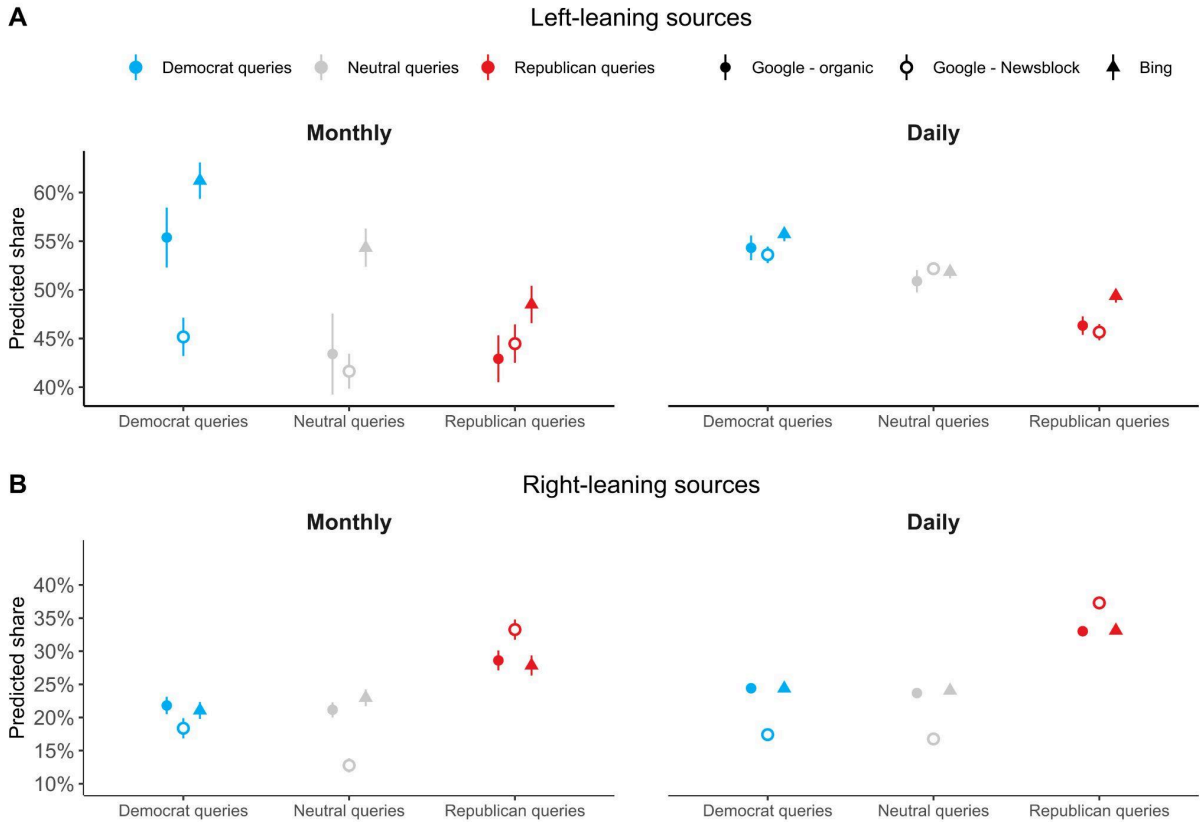

**Figure 2**. Predicted probabilities of media sources with a political leaning to the political left (Panel A) or right (Panel B) by query type (color) and search engine (symbol shape).

As expected in H1b, right-leaning media are consistently overrepresented for Republican queries both for the monthly and daily collections (see Panel B of Figure 2). Such sources were significantly more likely to be prioritized in response to Republican queries on Bing (monthly: *OR* = 1.49, *95%CI*[1.34:1.65], *p* < 0.001; daily: *OR* = 1.57, *95%CI*[1.51:1.62], *p* < 0.001 ) and Google organic search results (monthly: *OR* = 1.66, *95%CI*[1.51:1.83], *p* < 0.001; daily: *OR* = 1.55, *95%CI*[1.50:1.61], *p* < 0.001) as well as Google's Newsblock (monthly: *OR* = 3.13, *CI95%*[2.79:3.51], *p* < 0.001; daily: *OR* = 2.79, *95%CI*[2.69:2.89], *p* <0.001). Whereas Republican queries negatively predict the presence of left-leaning sources, the opposite pattern did not hold. Democratic queries never decrease the likelihood of right-leaning news sources in search outputs; in fact, they even increase the presence of right-leaning sources in monthly Google Newsblock searches (*OR* = 1.33, *95%CI*[1.16:1.53], *p* < 0.001).

## System-side factors of political information curation

So far we have established differences in the political leaning of media sources in search results depending on the user-side factors in the form of thematic focus of election-related queries: Democratic queries resulted in higher shares of left-leaning sources, while Republican queries increased the presence of right-leaning news sources. In the remainder of this section, we assess how these differences are influenced by two system-side factors, notably search localization and subsequent variation in search results based on the location from where the search has been conducted and the changes in search results over time due to evolving source relevance.

We first investigate to what extent the partisan gap in the political leaning of media sources is moderated by the search location. For the monthly collections, the difference in predicted probabilities of left-leaning sources between pro-Republican and pro-Democratic states only varied by a few percentage points (see Panel A of Figure 3). Specifically, we find no evidence that left-leaning sources are more prevalent in cases when the queries are conducted in a traditional 'blue' location (Los Angeles, CA) compared to a 'red' location (Dallas, TX). However, for the daily collections, we found instances where the gap in the predicted likelihood of left-leaning media surprisingly disappeared in 'blue' (Bing in Los Angeles, CA) or historically more mixed states (Bing in Columbus, OH; organic search results for Google in Iowa City, IA; organic search results for Google and Bing in Northern Virginia, VA). Note, however, that this regional variation remains within the range of a few percentage points.

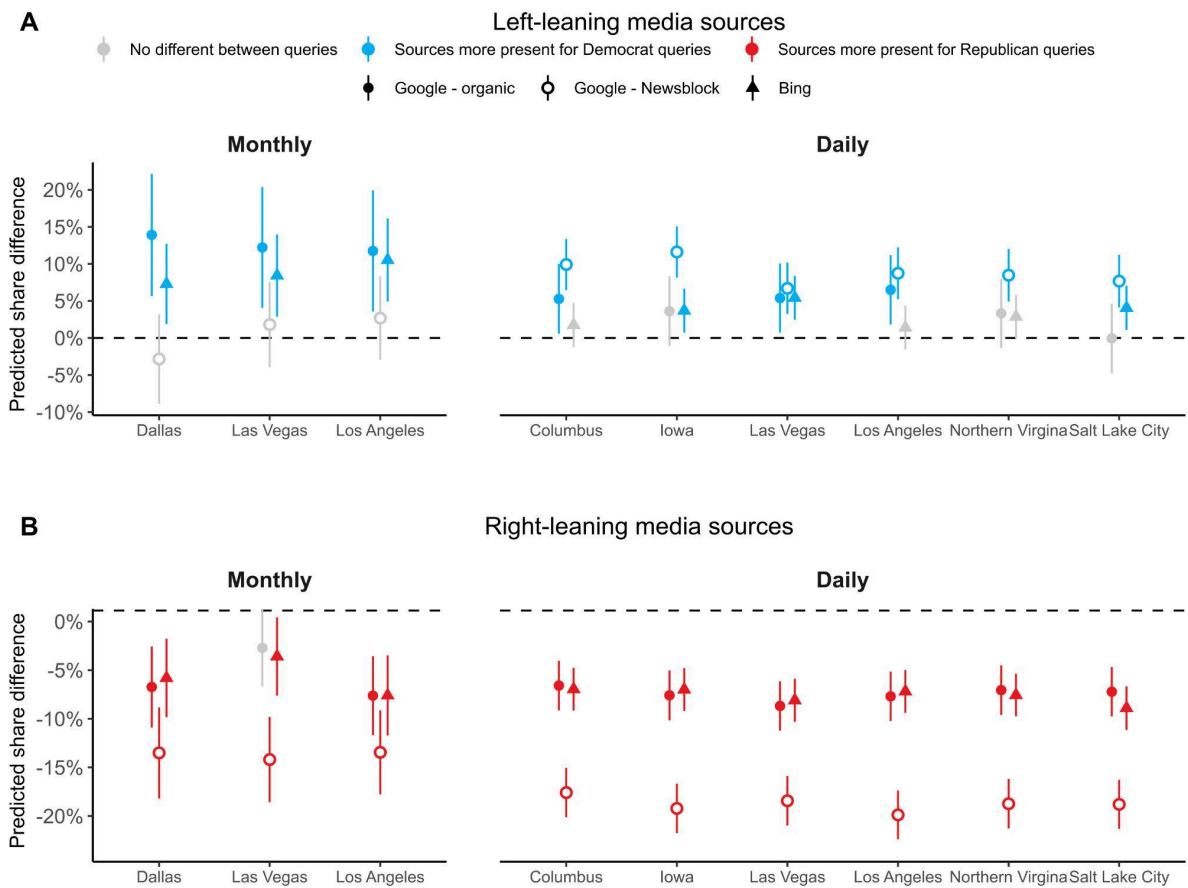

**Figure 3**. Difference in predicted probabilities of the likelihood of media sources with a leaning to the political left (Panel A) or right (Panel B) across locations (x-axis). Point estimates were calculated from Tukey-adjusted pairwise contrasts as the difference in predicted probability between Democratic and Republican queries. Positive (negative) estimates, therefore, indicate higher shares for Democratic (Republican) queries.

Panel B of Figure 3 illustrates the regional variation of the predicted difference in right-leaning news source prevalence. The results indicate that the increased likelihood of right-leaning news sources for Republican queries is remarkably consistent across geographical locations. The only exception is the monthly collections for Las Vegas, NV, where this positive association between right-leaning sources and Republican queries does not reach statistical significance.

Finally, we assess how the partisan gap in media source selection in search engine outputs changes over time. For the monthly collections, we find that the difference in left-leaning sources between Democratic and Republican queries tends to be smaller or absent in the first few months. For organic Google and Bing search results, the partisan gap in left-leaning news sources is largest in the later months of August through October. For the daily collections, the gap in left-leaning news sources tends to shrink in the final week before the election on 5 November 2024. While the extent of temporal variation is rather limited for organic Google and Bing searches, there is substantial short-term variation for Google Newsblock searches. Though Newsblock search results show higher shares of left-leaning sources for Democratic queries on most days, this pattern flips in the opposite direction in two instances. For instance, the difference in predicted shares is 13.5% (*95%CI*[8.03:19.1]) higher for Democratic queries on 3 November but then plummets to -17.8% (*95%CI*[-23.0:-12.6]) the next day, suggesting a significantly higher prevalence of left-leaning sources for Republican queries.

The inverse pattern of results emerges for right-leaning media. The relative overrepresentation of these media sources for Republican queries tends to be marginally smaller in the earlier months of the data collection but then stabilizes around a difference of roughly 10%. Once more the consistency of organic Google and Bing searches contrasts with stronger fluctuation in results for Google Newsblock.

1938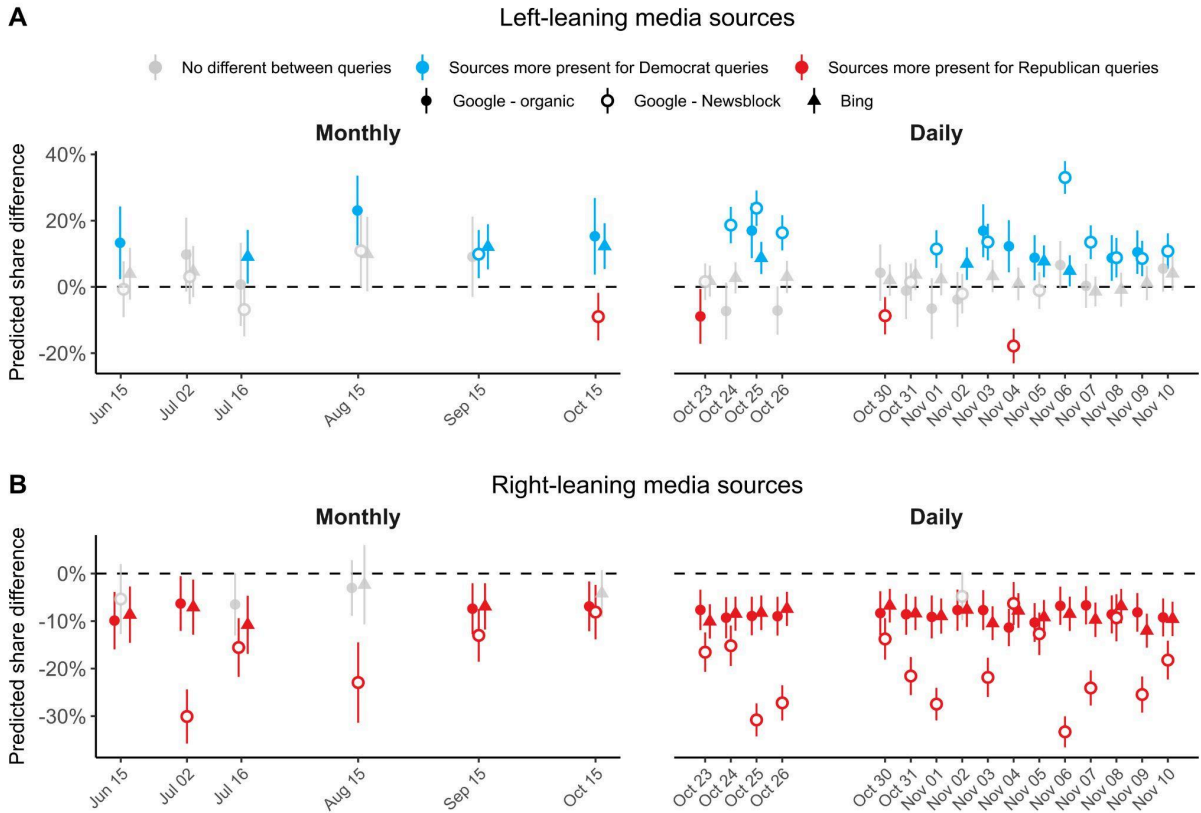

**Figure 4**. Difference in predicted probabilities of the likelihood of news sources with a leaning to the political left (Panel A) or right (Panel B) over time (x-axis). Point estimates were calculated from Tukey-adjusted pairwise contrasts as the difference in predicted probability between Democratic and Republican queries. Positive (negative) estimates, therefore, indicate higher shares for Democratic (Republican) queries.

# Discussion

In this article, we examined how the two largest Western search engines, Google and Bing, curated political information prior to the 2024 US elections. According to the search engines themselves, in particular, Google, their aim in this context was to "surface high-quality information to voters" (Jasper, 2023) by prioritizing authoritative local and regional news and information from the regional and state offices. Through our analysis, we aimed to understand not only whether this claim is empirically supported but also whether the curation of high-quality information by search engines was prone to (ideological) skewness and how it was affected by different user- and system-side factors.

One immediate finding of our study is that search engines tend to prioritize left-leaning media sources in relation to all types of election-related queries. While it does not mean that right-leaning media were completely excluded by search engines' curation systems, their presence was less pronounced, thus aligning to a certain degree with Republican claims about the skewed representation of electoral matters by search engines (e.g., Lowell, 2024). At the same time, it is important to note that the higher presence of left-leaning sources does not per se indicate that search engines presented a particular candidate in a more positive light, as some Republican accusations assumed. To verify such accusations, it would be necessary to conduct an analysis of specific content items linked by search engines, but such an analysis is beyond the scope of the current study. Furthermore, it has to be acknowledged that the partisan gap in media sources' prioritization can also be attributed to the multiple controversies associated with prominent right-leaning media's coverage of the 2020 elections that resulted in them being treated as less reliable sources by information curation systems and, consequently, downgraded in organic search results.

We also find support for our H1 that assumed that the ideological slant of search engine outputs aligns with the ideological slant of the search queries. The major implication of this finding is that search engines' information curation actively mirrors the partisan divides present in the US media environments - at least regarding the curation of media sources constituting the most common type of outputs on Google and Bing - and has the potential to contribute to (perceived) polarization within these environments. The exact degree to which such contribution is happening shall be measured in future research, but considering existing evidence of the substantive impact of search engines on opinion formation in the context of politics (Epstein & Robertson, 2015; Rohrbach et al., 2024) and the high trust towards search engines as an information source, we suggest that the risk of search engines contributing to societal polarization shall be treated seriously.

The above-mentioned observations also raise the question regarding the normative expectations about the role of search engines in the context of political communication in polarised media environments. With search engines serving as information gatekeepers that curate access to political information in a way that shares certain similarities with traditional forms of journalistic gatekeeping, the degree to which search engines' curation shall follow normative principles associated with journalism remains unclear. Shall, for instance, search engines reflect the (often biased) state of social reality at any given point in time, resulting in the unequal visibility of different ideologically slanted perspectives, which are the product of increasingly polarized media environments? Or shall search engines make interventions in line with the deliberative democracy principles (for some examples regarding other curation systems, see Helberger, 2019) to expose individuals to diverse information that can help negotiate different viewpoints and, hypothetically, counter societal polarization? The answers to these questions eventually depend on the desired role of information curation systems in democracies which currently remains unclear.

We also find that user-side factors turn out to be more impactful than system-side factors. In particular, for organic search results on Google and Bing, we observe relatively little variation over time and depending on. Interestingly, Google's Newsblock outputs turned out to be more volatile, which is likely due to the different algorithmic logic of curation used there. Such volatility can potentially counter the partisan divide mirrored in organic search results, albeit its higher customization can also lead to the opposite results. Overall, the additional search interfaces, such as the Newsblock, can play an important role in the context of (perceived) polarization, but the curation principles behind these interfaces are even less transparent than organic search. Under these circumstances, it is important to keep accounting for these additional forms of information curation, which are increasingly adopted by search engines in relation to specific issues and at specific points of time in future research.

Finally, it is important to note several limitations of the conducted study. First, to examine information curation, we relied just on a few search queries that referred to rather general election-related concepts and a few key actors. A relatively small pool of queries is attributed to the high amount of computational and human resources required to deploy and maintain infrastructure for data collection. However, it would be advantageous for future research to look for ways to expand the selection of queries used as well as to include more judgemental and ideologically slanted queries to assess how such a change in user-side factors will affect our observations regarding the performance of information curation systems.

Another limitation regards a relatively limited set of geographic locations, which we used to evaluate the effects of search localization on election-related information curation. Such a limitation is due to our reliance on Google Compute Engine for deploying infrastructure for the audit. Because of this reason, we had a limited set of US states for which we could apply IP addresses, including almost no "purple" (or battleground) states, where auditing information curation would be of particular interest. The last limitation regards our focus on analyzing political leaning only for media sources, which is both due to them being the most common type of source retrieved by search engines and the availability of data about their political leaning. For future research it may be advantageous to consider ways to expand the analysis of political leaning to other categories of sources, potentially by including the analysis of content linked in search engine results.

# Online Appendix

## Overview of sample

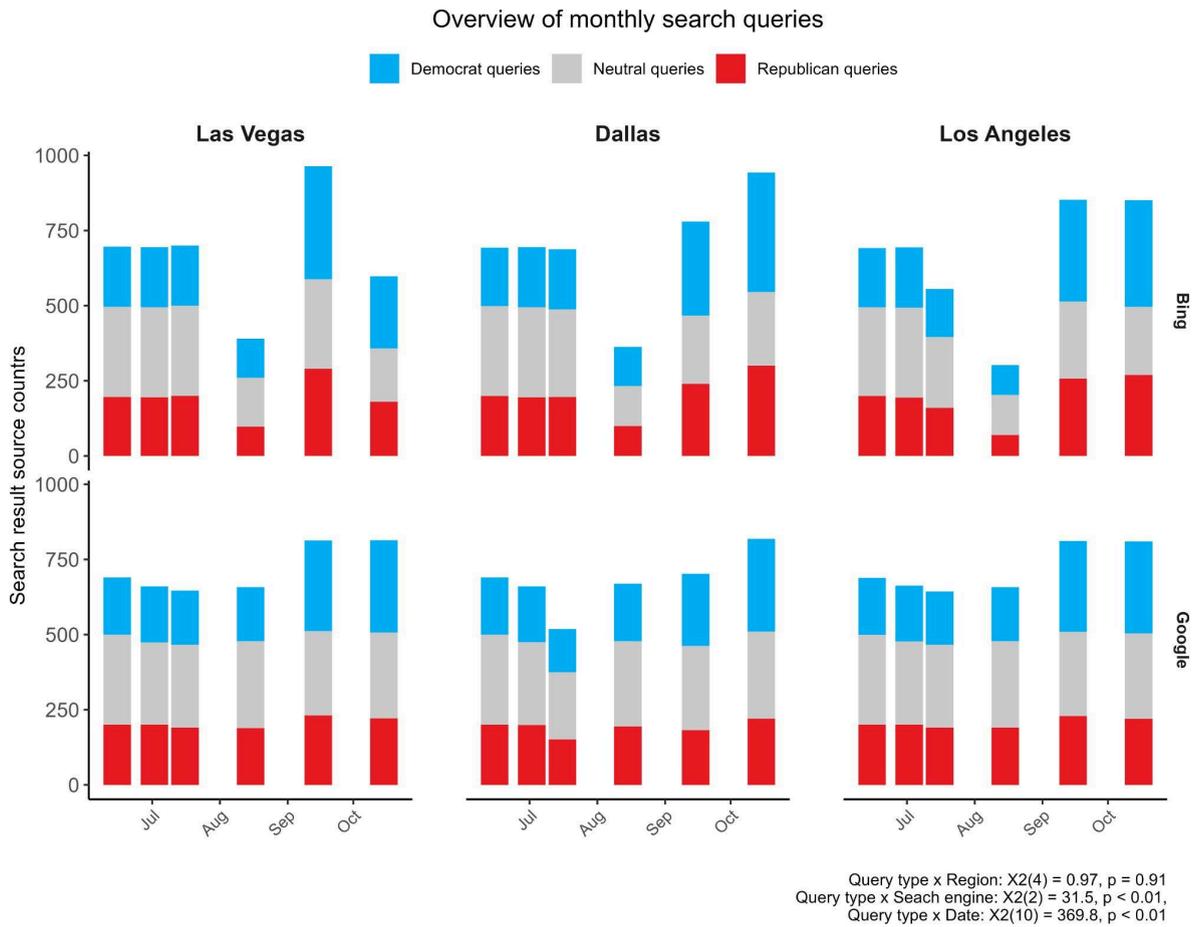

| | | **Google** | | | **Bing** | |
|---|---|---|---|---|---|---|
| | *Republican* | *Neutral* | *Democratic* | *Republican* | *Neutral* | *Democratic* |
| 2024-06-15 | 600 | 899 | 570 | 594 | 896 | 591 |
| 2024-07-02 | 599 | 827 | 557 | 584 | 900 | 600 |
| 2024-07-16 | 531 | 776 | 501 | 556 | 828 | 560 |
| 2024-08-15 | 573 | 861 | 551 | 267 | 429 | 360 |

| | | | | | |
|---|---|---|---|---|---|
| 2024-09-15 | 642 | 841 | 843 | 787 | 782 | 1027 |
| 2024-10-15 | 660 | 861 | 921 | 750 | 650 | 992 |
| Total | 3605 | 5065 | 3943 | 3538 | 4485 | 4130 |

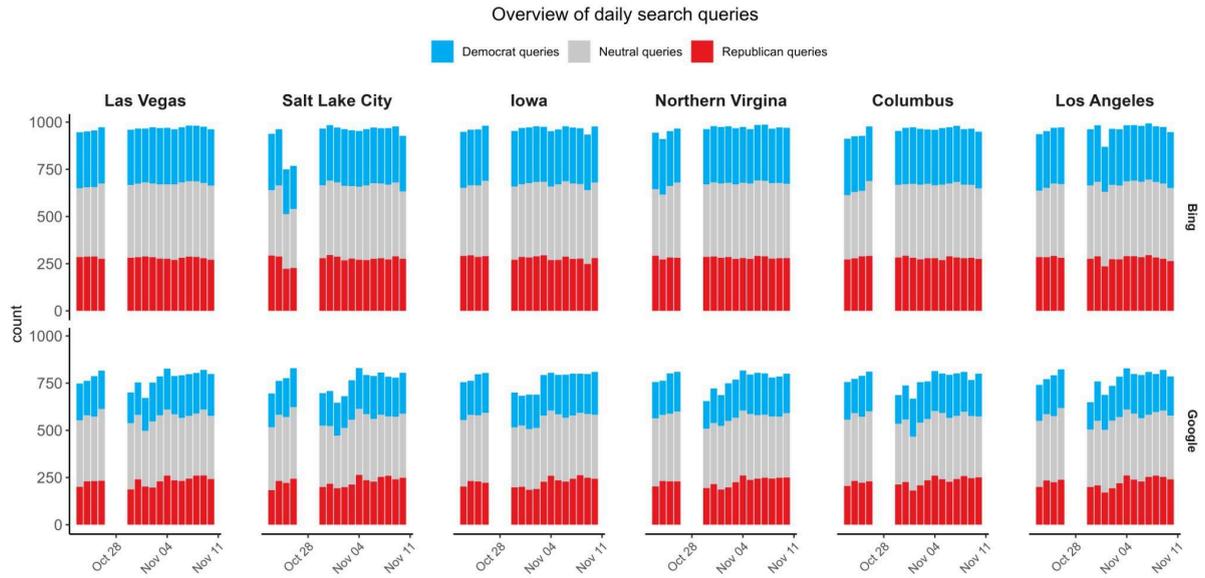

| | | **Google** | | | **Bing** | |
|---|---|---|---|---|---|---|
| | *Republican* | *Neutral* | *Democratic* | *Republican* | *Neutral* | *Democratic* |
| Before elections | 15864 | 24626 | 14174 | 20158 | 27609 | 21065 |
| After elections | 6000 | 8022 | 5140 | 6681 | 9449 | 7101 |
| Total | 21864 | 32648 | 19314 | 26839 | 37058 | 28166 |

# Additional tables

| Table B1 | | | | | | |
|---|---|---|---|---|---|---|
| | *Google queries* | | | *Bing queries* | | |
| Type of source | DEM | Neutral | REP | DEM | Neutral | REP |
| *Monthly searches* | | | | | | |
| journalistic | 25.3 | 10.7 | 44.7 | 71.7 | 61.2 | 82.0 |
| social media | 27.5 | 0 | 11.3 | 0.3 | 0 | 0 |
| government | 11.4 | 47.7 | 12.6 | 11.1 | 3.6 | 0 |
| candidate website | 6.0 | 3.3 | 9.7 | 0.1 | 0 | 2.5 |
| encyclopedia | 20.4 | 37.1 | 15.8 | 12.2 | 25.1 | 7.2 |
| NGO/think tank | 9.3 | 0.5 | 0 | 0.6 | 0.1 | 3.4 |
| research | 0 | 0.6 | 0 | 0.1 | 1.7 | 0 |
| blog | 0.1 | 0 | 2.4 | 3.9 | 8.3 | 8.3 |
| *Daily searches* | | | | | | |
| journalistic | 30.3 | 22.9 | 48.8 | 68.5 | 60.3 | 74.8 |
| social media | 26.0 | 0 | 11.9 | 0.2 | 0 | 0.1 |
| government | 15.8 | 39.4 | 11.7 | 10.1 | 2.7 | 0 |
| candidate website | 13.9 | 1.9 | 14.7 | 1.7 | 0 | 4.8 |
| encyclopedia | 13.2 | 32.8 | 11.8 | 13.7 | 22.3 | 14 |
| NGO/think tank | 0.6 | 0.5 | 1.0 | 1.7 | 0.3 | 0.2 |
| research | 0.2 | 2.4 | 0 | 0.4 | 4.9 | 0 |
| blog | 0 | 0 | 0.1 | 3.7 | 9.5 | 6.0 |

| Table B2.1 Monthly top 10 sources | | | | | | |
|---|---|---|---|---|---|---|
| Domain | Google - organic | | Google - Newsblock | | Bing | |
| | N | % | N | % | N | % |
| www.cnn.com | 823 | 26.2 | 663 | 8.44 | 1864 | 21.65 |
| apnews.com | 707 | 22.51 | 279 | 3.55 | 406 | 4.72 |
| www.economist.com | 489 | 15.57 | - | - | 138 | 1.6 |
| www.nytimes.com | 180 | 5.73 | 655 | 8.34 | 2565 | 29.79 |
| abcnews.go.com | 178 | 5.67 | - | - | - | - |
| www.usatoday.com | 94 | 2.99 | 440 | 5.6 | 217 | 2.52 |
| www.nbcnews.com | 80 | 2.55 | 181 | 2.31 | 1117 | 12.97 |
| www.washingtonpost.com | 74 | 2.36 | 197 | 2.51 | - | - |
| www.politico.com | 73 | 2.32 | 360 | 4.58 | 289 | 3.36 |
| www.cnbc.com | 63 | 2.01 | - | - | - | - |
| www.npr.org | - | - | - | - | 630 | 7.32 |
| www.bbc.com | - | - | - | - | 493 | 5.73 |
| www.reuters.com | - | - | - | - | 334 | 3.88 |
| www.theguardian.com | - | - | 595 | 7.58 | - | - |
| www.foxnews.com | - | - | 411 | 5.23 | - | - |
| www.aljazeera.com | - | - | 267 | 3.4 | - | - |
| Cumulative share of top 10 | | 87.91 | | 51.54 | | 93.54 |

| Table B2.2 Daily top 10 sources | | | | | | |
|---|---|---|---|---|---|---|
| Domain | Google - organic | | Google - Newsblock | | Bing | |
| | N | % | N | % | N | % |
| www.cnn.com | 4793 | 20.03 | 4777 | 6.72 | 10436 | 17.75 |
| apnews.com | 3885 | 16.23 | 2937 | 4.13 | 334 | 3.88 |
| www.nbcnews.com | 2038 | 8.51 | 5417 | 7.63 | 4549 | 7.74 |
| www.nytimes.com | 1796 | 7.5 | 5934 | 8.35 | 14192 | 24.13 |
| www.cbsnews.com | 1621 | 6.77 | - | - | - | - |
| www.reuters.com | 1028 | 4.29 | - | - | 2795 | 4.75 |
| www.bbc.com | 1012 | 4.23 | 2428 | 3.42 | 4424 | 7.52 |
| www.theguardian.com | 997 | 4.17 | 4338 | 6.11 | - | - |
| www.economist.com | 988 | 4.13 | - | - | 1569 | 2.67 |
| www.aljazeera.com | 890 | 3.72 | 3803 | 5.35 | - | - |
| www.usatoday.com | - | - | 3444 | 4.85 | 3049 | 5.18 |
| www.foxnews.com | - | - | 2945 | 4.15 | - | - |
| www.washingtonpost.com | - | - | 1778 | 2.5 | - | - |
| www.politico.com | - | - | - | - | 4147 | 7.05 |
| www.npr.org | - | - | - | - | 2757 | 4.69 |
| Cumulative share of top 10 | | 79.58 | | 53.21 | | 85.36 |

# Full regression tables

## Monthly data collection

| Table C1. Full regression results predicting left-leaning news sources in the monthly data collection | | | | | | | | | |
|---|---|---|---|---|---|---|---|---|---|
| | **Bing** | | | **Google - organic** | | | **Google - Newsblock** | | |
| (Intercept) | 1.05 | 0.93 – 1.20 | 0.426 | 0.92 | 0.72 – 1.17 | 0.499 | 0.72 | 0.63 – 0.83 | <0.001 |
| DEM query | 1.23 | 1.10 – 1.37 | <0.001 | 1.36 | 1.10 – 1.68 | 0.005 | 1.05 | 0.94 – 1.17 | 0.420 |
| REP query | 0.88 | 0.79 – 0.98 | 0.022 | 0.86 | 0.70 – 1.04 | 0.123 | 0.99 | 0.89 – 1.11 | 0.880 |
| Dallas, TX | 1.07 | 0.96 – 1.19 | 0.196 | 1.12 | 0.94 – 1.34 | 0.195 | 1.03 | 0.92 – 1.15 | 0.658 |
| L.A., CA | 1.07 | 0.96 – 1.19 | 0.204 | 0.91 | 0.77 – 1.08 | 0.300 | 1.01 | 0.91 – 1.13 | 0.852 |
| 02.07.2024 | 1.21 | 1.05 – 1.40 | 0.008 | 0.89 | 0.70 – 1.12 | 0.320 | 0.97 | 0.83 – 1.13 | 0.684 |
| 16.07.2024 | 1.19 | 1.03 – 1.38 | 0.021 | 0.83 | 0.65 – 1.07 | 0.147 | 1.10 | 0.94 – 1.28 | 0.242 |
| 15.08.2024 | 0.95 | 0.80 – 1.14 | 0.593 | 0.57 | 0.45 – 0.73 | <0.001 | 1.12 | 0.94 – 1.34 | 0.197 |
| 15.09.2024 | 1.07 | 0.93 – 1.23 | 0.364 | 0.93 | 0.73 – 1.19 | 0.565 | 1.12 | 0.97 – 1.29 | 0.133 |
| 15.10.2024 | 1.02 | 0.89 – 1.18 | 0.760 | 1.22 | 0.96 – 1.55 | 0.105 | 0.97 | 0.83 – 1.12 | 0.649 |
| **Random Effects** | | | | | | | | | |

| | | | | | | | |
|---|---|---|---|---|---|---|---|
| $\sigma^2$ | 3.29 | | 3.29 | | 3.29 | | |
| $\tau_{00}$ | 0.01 $_{html\_name}$ | | 0.01 $_{html\_name}$ | | 0.00 $_{html\_name}$ | | |
| ICC | 0.00 | | 0.00 | | 0.00 | | |
| N | 1254 $_{html\_name}$ | | 1254 $_{html\_name}$ | | 1380 $_{html\_name}$ | | |
| Obs. | 8583 | | 8583 | | 7783 | | |
| $R^2$ /Cond.$R^2$ | 0.008 / 0.012 | | 0.008 / 0.012 | | 0.001 / 0.001 | | |

| Table C2. Full regression results predicting right-leaning news sources in the monthly data collection | | | | | | | | | |
|---|---|---|---|---|---|---|---|---|---|
| | **Bing** | | | **Google - organic** | | | **Google - Newsblock** | | |
| (Intercept) | 0.27 | 0.23 – 0.30 | <0.001 | 0.29 | 0.26 – 0.33 | <0.001 | 0.13 | 0.11 – 0.15 | <0.001 |
| DEM query | 1.02 | 0.92 – 1.13 | 0.707 | 1.03 | 0.93 – 1.14 | 0.589 | 1.33 | 1.16 – 1.53 | <0.001 |
| REP query | 1.49 | 1.34 – 1.65 | <0.001 | 1.66 | 1.51 – 1.83 | <0.001 | 3.13 | 2.79 – 3.51 | <0.001 |
| Dallas, TX | 0.97 | 0.88 – 1.08 | 0.576 | 1.00 | 0.90 – 1.10 | 0.937 | 1.10 | 0.97 – 1.23 | 0.132 |
| L.A., CA | 1.00 | 0.91 – 1.11 | 0.940 | 0.95 | 0.86 – 1.05 | 0.318 | 1.01 | 0.90 – 1.14 | 0.834 |
| 02.07.2024 | 1.17 | 1.01 – 1.35 | 0.030 | 1.00 | 0.87 – 1.15 | 0.991 | 1.36 | 1.14 – 1.63 | 0.001 |
| 16.07.2024 | 1.02 | 0.88 – 1.18 | 0.790 | 0.97 | 0.83 – 1.12 | 0.650 | 1.29 | 1.09 – 1.54 | 0.004 |
| 15.08.2024 | 1.02 | 0.86 – 1.22 | 0.820 | 0.89 | 0.77 – 1.03 | 0.109 | 1.06 | 0.86 – 1.31 | 0.581 |
| 15.09.2024 | 1.08 | 0.94 – 1.24 | 0.273 | 0.96 | 0.84 – 1.10 | 0.587 | 1.26 | 1.07 – 1.49 | 0.006 |
| 15.10.2024 | 1.05 | 0.91 – 1.21 | 0.500 | 0.85 | 0.74 – 0.98 | 0.024 | 1.45 | 1.23 – 1.70 | <0.001 |
| **Random Effects** | | | | | | | | | |
| $\sigma^2$ | 3.29 | | | 3.29 | | | 3.29 | | |
| $\tau_{00}$ | 0.00 $_{html\_name}$ | | | 0.01 $_{html\_name}$ | | | 0.00 $_{html\_name}$ | | |
| ICC | 0.00 | | | 0.00 | | | 0.00 | | |
| N | 1254 $_{html\_name}$ | | | 1412 $_{html\_name}$ | | | 1380 $_{html\_name}$ | | |
| Obs. | 12153 | | | 12613 | | | 9857 | | |
| $R^2$ /Cond.$R^2$ | 0.010 / 0.011 | | | 0.016 / 0.017 | | | 0.001 / 0.016 | | |

Daily data collection

| Table C3. Full regression results predicting left-leaning news sources in the daily data collection | | | |
|---|---|---|---|
| | **Bing** | **Google - organic** | **Google - Newsblock** |

|  |  |  |  |  |  |  |  |  |  |
|---|---|---|---|---|---|---|---|---|---|
| (Intercept) | 1.13 | 1.04 – 1.22 | 0.002 | 1.41 | 1.23 – 1.62 | <0.001 | 1.19 | 1.10 – 1.29 | <0.001 |
| DEM query | 1.17 | 1.12 – 1.22 | <0.001 | 0.96 | 0.90 – 1.03 | 0.264 | 1.04 | 1.00 – 1.08 | 0.063 |
| REP query | 0.94 | 0.90 – 0.97 | 0.001 | 0.73 | 0.69 – 0.78 | <0.001 | 0.77 | 0.74 – 0.80 | <0.001 |
| Salt Lake City, UT | 1.02 | 0.96 – 1.07 | 0.597 | 0.96 | 0.88 – 1.05 | 0.367 | 1.02 | 0.97 – 1.08 | 0.413 |
| Iowa City, IA | 1.00 | 0.95 – 1.06 | 0.995 | 0.99 | 0.90 – 1.08 | 0.738 | 0.93 | 0.88 – 0.99 | 0.016 |
| Northern Virginia, VA | 1.02 | 0.97 – 1.08 | 0.465 | 0.98 | 0.90 – 1.07 | 0.661 | 1.01 | 0.95 – 1.06 | 0.851 |
| Columbus, OH | 1.02 | 0.96 – 1.08 | 0.506 | 0.99 | 0.90 – 1.08 | 0.755 | 0.92 | 0.87 – 0.98 | 0.005 |
| L.A., CA | 1.02 | 0.96 – 1.07 | 0.601 | 0.97 | 0.89 – 1.06 | 0.482 | 1.00 | 0.94 – 1.05 | 0.914 |
| 2024-10-24 | 1.06 | 0.96 – 1.16 | 0.246 | 0.84 | 0.71 – 0.99 | 0.032 | 1.13 | 1.02 – 1.24 | 0.018 |
| 2024-10-25 | 0.93 | 0.85 – 1.02 | 0.127 | 0.95 | 0.81 – 1.12 | 0.581 | 1.09 | 0.98 – 1.20 | 0.103 |
| 2024-10-26 | 0.98 | 0.90 – 1.08 | 0.718 | 0.75 | 0.65 – 0.88 | <0.001 | 0.96 | 0.87 – 1.06 | 0.461 |
| 2024-10-30 | 0.98 | 0.89 – 1.07 | 0.626 | 0.89 | 0.75 – 1.04 | 0.145 | 0.76 | 0.69 – 0.84 | <0.001 |
| 2024-10-31 | 0.96 | 0.87 – 1.05 | 0.344 | 0.66 | 0.56 – 0.78 | <0.001 | 0.72 | 0.65 – 0.80 | <0.001 |
| 2024-11-01 | 0.95 | 0.86 – 1.04 | 0.254 | 0.82 | 0.69 – 0.98 | 0.026 | 1.08 | 0.98 – 1.19 | 0.115 |
| 2024-11-02 | 0.97 | 0.88 – 1.06 | 0.483 | 0.67 | 0.57 – 0.79 | <0.001 | 0.97 | 0.88 – 1.07 | 0.504 |
| 2024-11-03 | 0.93 | 0.85 – 1.02 | 0.134 | 0.80 | 0.68 – 0.94 | 0.006 | 0.93 | 0.85 – 1.02 | 0.143 |
| 2024-11-04 | 0.94 | 0.85 – 1.03 | 0.158 | 0.80 | 0.68 – 0.93 | 0.004 | 0.79 | 0.72 – 0.87 | <0.001 |
| 2024-11-05 | 0.87 | 0.80 – 0.96 | 0.003 | 1.07 | 0.92 – 1.24 | 0.410 | 0.91 | 0.82 – 1.00 | 0.045 |
| 2024-11-06 | 0.91 | 0.84 – 1.00 | 0.050 | 0.89 | 0.76 – 1.03 | 0.107 | 0.83 | 0.76 – 0.91 | <0.001 |
| 2024-11-07 | 0.97 | 0.89 – 1.07 | 0.573 | 0.98 | 0.85 – 1.14 | 0.800 | 1.02 | 0.94 – 1.12 | 0.600 |
| 2024-11-08 | 0.85 | 0.78 – 0.94 | 0.001 | 0.78 | 0.68 – 0.90 | 0.001 | 1.00 | 0.91 – 1.09 | 0.972 |
| 2024-11-09 | 0.91 | 0.82 – 0.99 | 0.038 | 0.81 | 0.70 – 0.94 | 0.005 | 1.14 | 1.04 – 1.25 | 0.004 |
| 2024-11-10 | 0.90 | 0.82 – 0.99 | 0.027 | 0.80 | 0.69 – 0.92 | 0.002 | 1.09 | 0.99 – 1.19 | 0.085 |
| **Random Effects** | | | | | | | | | |
| $\sigma^2$ | 3.29 | | | 3.29 | | | 3.29 | | |
| $\tau_{00}$ | 0.00 html_name | | | 0.01 html_name | | | 0.06 html_name | | |
| ICC | 0.00 | | | 0.00 | | | 0.02 | | |

| N | 9106 html_name | | 8014 html_name | | 9591 html_name | |
|---|---|---|---|---|---|---|
| Obs. | 58746 | | 23935 | | 70479 | |
| $R^2$ /Cond.$R^2$ | 0.003 / 0.003 | | 0.011 / 0.011 | | 0.010 / 0.028 | |

| Table C4. Full regression results predicting right-leaning news sources in the daily data collection | | | | | | | | | |
|---|---|---|---|---|---|---|---|---|---|
| | **Bing** | | | **Google - organic** | | | **Google - Newsblock** | | |
| (Intercept) | 0.32 | 0.30 – 0.34 | <0.001 | 0.30 | 0.28 – 0.33 | <0.001 | 0.23 | 0.21 – 0.25 | <0.001 |
| DEM query | 0.98 | 0.94 – 1.02 | 0.281 | 0.99 | 0.95 – 1.03 | 0.731 | 1.02 | 0.96 – 1.07 | 0.563 |
| REP query | 1.57 | 1.51 – 1.62 | <0.001 | 1.55 | 1.50 – 1.61 | <0.001 | 2.79 | 2.69 – 2.89 | <0.001 |
| Salt Lake City, UT | 0.95 | 0.90 – 1.00 | 0.047 | 0.94 | 0.89 – 1.00 | 0.046 | 0.98 | 0.93 – 1.04 | 0.521 |
| Iowa City, IA | 0.99 | 0.94 – 1.04 | 0.644 | 0.98 | 0.93 – 1.04 | 0.556 | 0.99 | 0.94 – 1.05 | 0.772 |
| Northern Virginia, VA | 0.98 | 0.93 – 1.03 | 0.408 | 0.97 | 0.92 – 1.03 | 0.318 | 1.00 | 0.95 – 1.06 | 0.901 |
| Columbus, OH | 0.99 | 0.94 – 1.05 | 0.808 | 0.98 | 0.93 – 1.04 | 0.460 | 0.99 | 0.94 – 1.05 | 0.765 |
| L.A., CA | 1.02 | 0.97 – 1.07 | 0.455 | 0.98 | 0.93 – 1.04 | 0.539 | 1.00 | 0.95 – 1.06 | 0.909 |
| 2024-10-24 | 0.96 | 0.88 – 1.05 | 0.376 | 1.02 | 0.93 – 1.13 | 0.610 | 0.97 | 0.88 – 1.07 | 0.534 |
| 2024-10-25 | 0.96 | 0.88 – 1.05 | 0.383 | 1.07 | 0.97 – 1.18 | 0.158 | 0.83 | 0.75 – 0.91 | <0.001 |
| 2024-10-26 | 0.95 | 0.88 – 1.04 | 0.276 | 1.06 | 0.96 – 1.16 | 0.242 | 0.89 | 0.81 – 0.99 | 0.025 |
| 2024-10-30 | 1.04 | 0.96 – 1.13 | 0.324 | 1.09 | 0.99 – 1.21 | 0.067 | 0.92 | 0.83 – 1.01 | 0.073 |
| 2024-10-31 | 0.99 | 0.91 – 1.08 | 0.838 | 1.11 | 1.01 – 1.22 | 0.033 | 0.88 | 0.80 – 0.97 | 0.010 |
| 2024-11-01 | 0.97 | 0.89 – 1.06 | 0.521 | 1.10 | 1.00 – 1.21 | 0.053 | 0.82 | 0.74 – 0.91 | <0.001 |
| 2024-11-02 | 1.04 | 0.95 – 1.13 | 0.390 | 1.06 | 0.97 – 1.17 | 0.199 | 1.02 | 0.92 – 1.12 | 0.729 |
| 2024-11-03 | 0.96 | 0.88 – 1.04 | 0.342 | 1.04 | 0.94 – 1.14 | 0.441 | 0.87 | 0.79 – 0.96 | 0.005 |
| 2024-11-04 | 0.98 | 0.90 – 1.06 | 0.578 | 1.03 | 0.94 – 1.13 | 0.549 | 1.05 | 0.95 – 1.15 | 0.356 |
| 2024-11-05 | 1.01 | 0.93 – 1.10 | 0.863 | 1.00 | 0.91 – 1.10 | 0.958 | 0.99 | 0.90 – 1.08 | 0.772 |
| 2024-11-06 | 1.02 | 0.94 – 1.11 | 0.658 | 1.09 | 0.99 – 1.19 | 0.084 | 0.81 | 0.74 – 0.89 | <0.001 |
| 2024-11-07 | 0.97 | 0.90 – 1.06 | 0.541 | 1.07 | 0.98 – 1.18 | 0.130 | 0.90 | 0.82 – 0.98 | 0.019 |
| 2024-11-08 | 0.99 | 0.91 – 1.08 | 0.825 | 1.15 | 1.05 – 1.26 | 0.003 | 0.94 | 0.86 – 1.03 | 0.192 |
| 2024-11-09 | 0.97 | 0.89 – 1.05 | 0.438 | 1.09 | 0.99 – 1.20 | 0.065 | 0.86 | 0.78 – 0.94 | 0.001 |

| 2024-11-10 | 0.99 | 0.91 – 1.08 | 0.808 | 1.08 | 0.99 – 1.19 | 0.090 | 0.88 | 0.80 – 0.97 | 0.010 |
|---|---|---|---|---|---|---|---|---|---|
| **Random Effects** | | | | | | | | | |
| $\sigma^2$ | 3.29 | | | 3.29 | | | 3.29 | | |
| $\tau_{00}$ | 0.00 $_{html\_name}$ | | | 0.01 $_{html\_name}$ | | | 0.04 $_{html\_name}$ | | |
| ICC | 0.00 | | | 0.00 | | | 0.01 | | |
| N | 9545 $_{html\_name}$ | | | 9244 $_{html\_name}$ | | | 9592 $_{html\_name}$ | | |
| Obs. | 92063 | | | 73826 | | | 87384 | | |
| $R^2$ /Cond.$R^2$ | 0.013 / 0.013 | | | 0.013 / 0.013 | | | 0.065 / 0.074 | | |